Adam JANIAK\*, Damian KOWALCZYK\*, Maciej LICHTENSTEIN\*

# PARALLEL/DISTRIBUTED TABU SEARCH FOR SCHEDULING MICROPROCESSOR TASKS IN HYBRID FLOW SHOP

The paper deals with the makespan minimization in the hybrid flow shop scheduling problem with multiprocessor tasks. The hybrid flow shop (HFS) generalizes the classical flow shop processor configuration by replacing each processor (processing stage) by some number of identical parallel processors. Similarly, the multiprocessor tasks generalize the classical assumption, by allowing a task to require more than one processor simultaneously for its processing. In this work we present the algorithm for solving the problem based on the tabu search technique. The proposed algorithm uses parallel and distributed mechanisms for neighborhood evaluation and well balances heterogeneous network environment.

## 1. INTRODUCTION

The problem of scheduling multiprocessor tasks in hybrid flow shop (HFS) processing environment has attracted many researchers all over the world. This is caused mainly by its good practical motivation, since the HFS models many real-life production systems and its applications can be also found in computer systems (e.g. computer vision)[1]. Since the considered problem is strongly NP-hard, the efficient (polynomial) algorithms for solving it are no likely to exist. Thus, the research on this problem focused mainly on developing heuristic [1] and metaheuristic [2,3] algorithms for finding "good" solutions in reasonable time. However, the algorithms available in the literature are capable to solve only moderate-size instances of the problem on the modern PC. This motivated us to develop tabu-search-based algorithm for solving the problem that utilizes the power of multicore processors and can be distributed in heterogeneous network.

## 2. PROBLEM FORMULATION

The problem considered in this paper can be stated as follows. The HFS is defined by the set $M=\{1,\ldots,m\}$ of $m$ processing stages. At each stage $i \in M$ there is a set $M_i=\{1,\ldots,m_i\}$ of $m_i$ identical processors. The set $J=\{1,\ldots,n\}$ of n multiprocessor jobs has to be processed in HFS defined by the sets $M$ and $M_i$, $i \in M$. Each job $j \in J$ is comprised of the set $\{o_{1j}, o_{2j},\ldots,o_{mj}\}$ of $m$ multiprocessor tasks. Each task $o_{ij}$, $i \in M$, $j \in J$ requires $size_{ij}$ ($1 \leq size_{ij} \leq m_i$) processors of the $i$-th stage and occupies them for $p_{ij}>0$ time units for its processing. Additionally, the processing of task $o_{ij}$, cannot be started before the processing of task $o_{(i-1)j}$ competes. Let $C_{ij}(s)$ denote the completion time of task $o_{ij}$ in some schedule $s$. The problem is to find the schedule $s^*$ such that:
$$C_{max}(s^*) = \min_s \{C_{max}(s)\}$$
where $C_{max}(s) = \max_{i \in M, j \in J} \{C_{ij}(s)\}$ is the makespan of schedule $s$.

---

\* Institute of Computer Science, Control and Robotics, Wrocław University of Technology, Janiszewskiego 11/17 50372 Wrocław, Poland, e-mail: {adam.janiak, damian.kowalczyk, maciej.lichtenstein}@pwr.wroc.pl

## 3. TABU SEARCH

The developed tabu-search-based algorithm is basically the same as the one presented in [2]. Since more than 95% of tabu search running time is consumed by the neighbourhood evaluation, in our implementation we distribute the computation over heterogeneous computer network. Our implementation is a single client – multiple server application, where each server evaluate the part of the neighbourhood commissioned by the single client. Every server performs its calculation by running parallel processes to utilize all available CPUs (or its cores). The main strength of our approach is the load balancing methods. The main goal of the load balancer is to minimize the idle time of each server that can appear between consecutive iterations of tabu search. To achieve this, we introduced the hierarchical method of grouping servers into nodes (super servers), where each node can be seen by the client as a single server, and the client do not have to care of the distribution of the computations within a node. Moreover, every server within a node can be another super server, so the load balancing within the node and at the client is limited to the small number of servers. We also introduce some methods of dynamic load balancing that take into account variable conditions in the network (even variable number of servers over time).

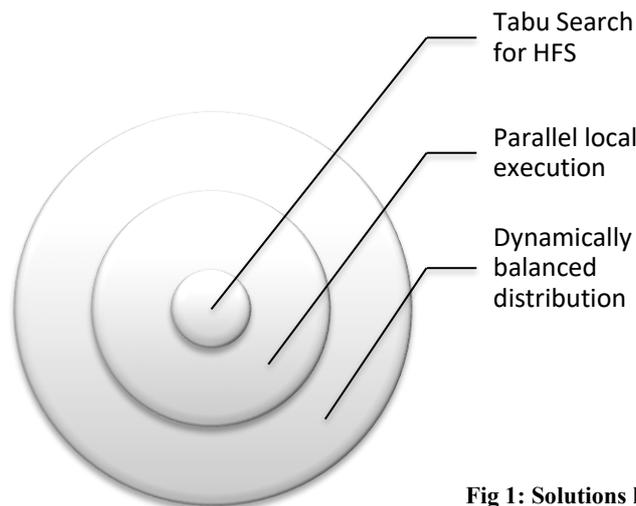

**Fig 1: Solutions hierarchy**

## 4. PARALLEL TS FOR HFS

First step towards performance and scalability - in a local homogenous environment, consists of parallelizing the implemented TS solution. In regard to findings in [2] and [4] most of the efforts should and will revolve around the process of local neighbourhood evaluation. The aim of this step is to gain as much performance as possible from the available processing units in the system.

The way to achieve that can be brought down to dividing the evaluation process into separate threads. They would be later assigned by the *task scheduler* of the operating system to available logical units (normally physical processors), to utilize full computing power of the host. Parallelizing that process is not a trivial task and is strictly dependent on the algorithm's dynamics.

Important virtue of Tabu Search is its potential for parallel execution. That is due to the character of a single iteration, which consists of substantial number of evaluations, all of them being mutually independent. That allows us to execute them simultaneously in groups of own choice. In order to avoid redundancy of efforts, each concurrent evaluation is focused on a separate part of the entire local neighbourhood (for a current solution / iteration). The best move (leading to a local minimum) is then selected among best moves established in each of partial evaluations. Draft of this approach has first been presented in [4].

For the above purpose a stand-alone component has been designed to encapsulate the parallelized evaluation, carrying out number of technical tasks:

➢ Neighbourhood partitioning

A technique has been implemented to divide a solution's neighborhood in *O(1)* time. Furthermore, it has been assumed for a local host, that performance of every processing unit is equal. In natural consequence, the local neighbourhood is divided into equal parts, to minimize the computing time of any iteration.

➢ Neighbourhood evaluation

Once the component is initialized with a HFS *schedule* and an (initially empty) *tabu list* of forbidden moves – both founding contextual data for the process, search for the local minimum can start. Given neighbourhood is being partitioned, and every part is being processed in a separate thread. Best move is chosen among partial results.

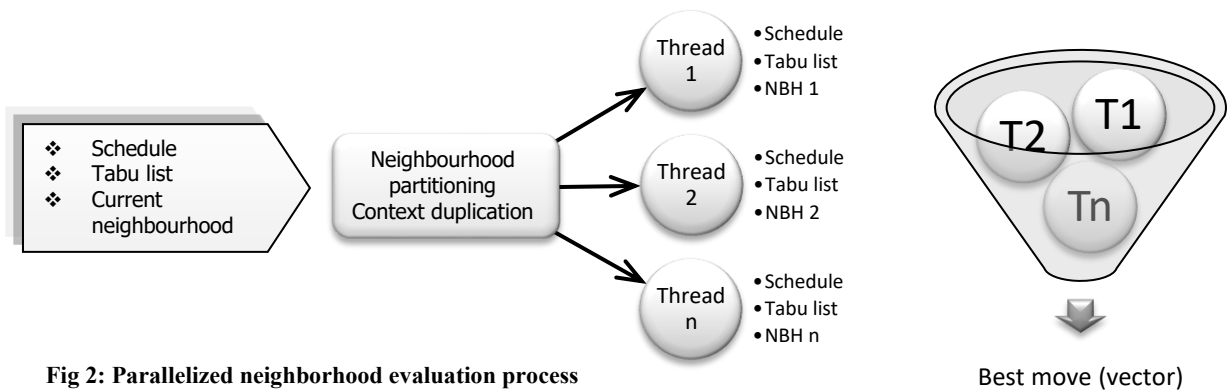

**Fig 2: Parallelized neighborhood evaluation process**

➢ Thread synchronization

To maximize performance, processing context is being duplicated for every thread, in order for all to work on a physically separate set of data. That minimizes the duration of delays caused by synchronization events.

➢ Event handling

All threads upon completion raise result events to the component, which after collecting the last result selects the best move (leading to the local minimum) and returns it with a global event. Among other events generated by the partial evaluations, worth mentioning are progress and error reporting, designed with regard to reliability and responsiveness of the process. All of them are handled either internally, or by external listeners of the component. This functionality will be heavily relied upon, in the next stage of the project.

## 5. SYNTHIA - DISTRIBUTED SOLVER

Once the local parallelization of TS has been achieved and encapsulated, we can reach beyond the performance of one host, and distribute the optimization process over computer network. Ideally an end user of the solver should not be concerned if the process is parallelized among a number of processors and if they are distributed geographically. Distribution transparency and scalability become two of our top goals in designing a distributed solver.

Heterogeneous network environment introduces number of variables the system has to tackle. Not only, by definition, will performance of most computing nodes and network links be different, but also variable in time due to random usage patterns. Failures (periodic time out of service) of both are another challenge. To address all of them, dynamic load balancing mechanisms are implemented.

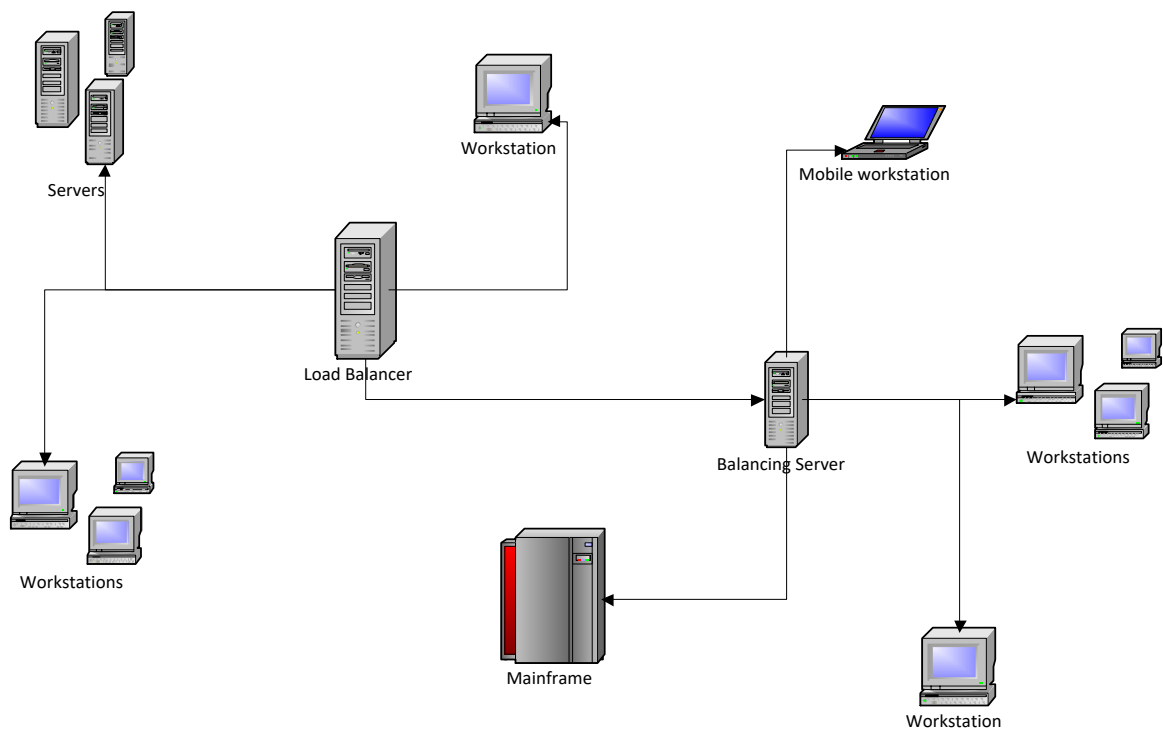

Fig 3: Example of heterogeneous network environment with balancing

## 5.1 Two distinct entities

The design of the solver assumes centralized service-oriented architecture (SOA) for efficient management and balancing of the optimization process. (*Why* loosely coupled) Two entities are introduced in consequence.

a) Service provider

The sole purpose of this role is to expose a web service wrapping the parallel TS component described in previous chapter, and therefore providing parallelization transparency to remote client. This role is performed by every computing node.

b) Load balancer / service consumer

The duty of this entity is to manage the optimization across a given set of distributed computing nodes. This includes dynamic adaptation of the process, to network state and parameters, monitoring of results and progress, but also handling of host / network failures. The measure of success for the balancer would be how busy it can keep all the nodes, how quickly it recovers from any node's failure, and eventually how soon it can deliver the result of last iteration.

There is virtually no obstacle for a balancer with subordinate nodes, to be treated as a standalone computing node, serving a combined computing potential in a wider distribution of the process. In such case a host will perform duties of both entities (refer to Fig. 3).

## 5.2 Performance prediction

Overall performance of the optimization depends on how well the solution is distributed across all nodes in subsequent iterations. Ideally all partial evaluations are finished in the same moment, and a next series can start with no delay. We attempt to achieve that by dividing the solution's neighborhood proportionally to each node's performance (expressed by number of schedule vectors processed per second)

$$n_i = \frac{p_i}{P} \times N$$

where: $p_i$ – performance of node $i$, $n_i$ – size of a neighborhood to be processed by $i$, $P$ – summary performance of the system, $N$ – size of entire neighborhood

Given a fully unloaded network setting, and machines dedicated to our optimization task only, this mechanism could turn out quite reliable for every iteration. Real life examples in heterogeneous environment though, introduce external network traffic and workload, resulting in performance fluctuations, which have to be covered dynamically.

Our prediction mechanisms depend on statistics, attempting to infer future performance of every node (overlapping its network performance) from historical values gathered after each iteration.

The method we used regularly, takes a weighted average from past performance results, where a weight of a particular measurement is equal to the size of a solution's neighborhood processed:

$$p_{ij} = \frac{\sum_{k=0}^{j-1} n_k p_k}{\sum_{k=0}^{j-1} n_k}, \text{ where } j - \text{iteration number}$$

We also introduce four other prediction methods, which currently await further testing.

### 5.3 Load balancing

Optimization process begins with performance assessment of all assigned computing nodes. Evaluation task is sent out for a generated mid-complexity schedule, and all partial processes are stopped after a certain period of time. All nodes return a (possibly incomplete) result, together with a statistical average speed measured. Those measurements, initializing the performance history of each node, allow the balancer to start distributing our Tabu Search solution proportionally to forecasts described in 5.2. Subsequent iterations are finished with receiving the last partial result, unless a timeout has occurred. In case of an incomplete iteration, the remaining evaluation is repeated in an extra iteration, before the process continues. Remaining neighborhood, depending on its size, is also distributed and remotely parallelized, to minimize processing time.

After a successful iteration, tabu list is updated along with the criterion value, and current schedule solution translated with the best move (vector) chosen among all partial evaluations of last iteration. If the initially specified number of iterations has been performed, solver rests, returning the best schedule discovered.

Additionally a diversification mechanism has been implemented to take the optimization into new regions. After a specified number of consequent unsuccessful iterations (resulting in the schedule translation to lesser criterion) a randomization is performed by applying n random vectors to the current solution, before the optimization continues.

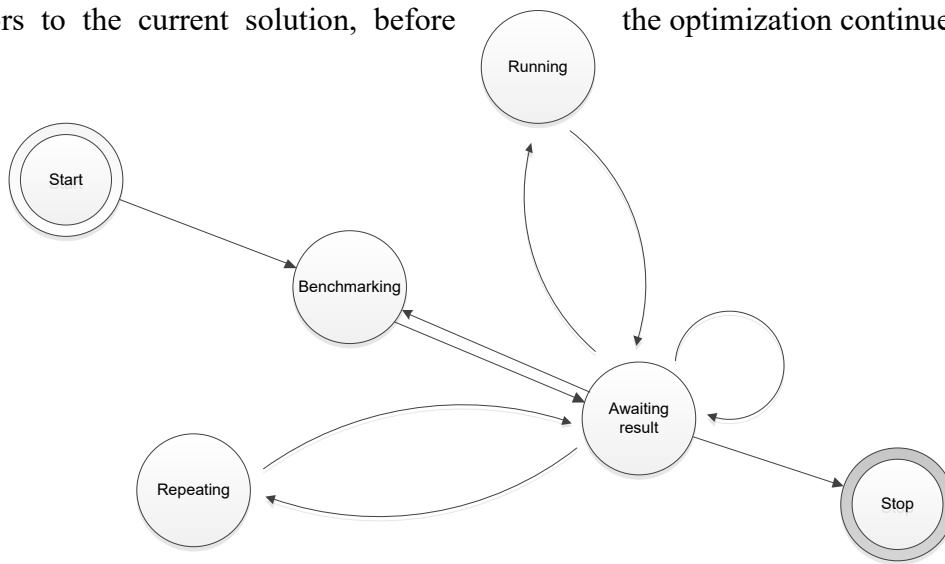

**Fig 4: Load balancing in heterogeneous environment**

### 5.4 Fault tolerance

Performance of the optimization is heavily dependent on reliability of the system. It is crucial for the load balancer to recover from any remote failures as soon as possible. In respect of that, mechanisms have been implemented on the web service and balancer side, for state monitoring and handling, to minimize any delays affecting the optimization. Remote node, on exit reports its state to the balancer, so that its results or cooperation are not expected in the next iteration. Network failures are covered by the balancer, exercising timed assertions for communication channels. All remote nodes are handled and monitored by separate local proxies, executed asynchronously. Those techniques proved invaluable during our tests in heterogeneous setting.

## 6. Experiments

Series of tests have been conducted to estimate the scalability and performance of the solution.

### 6.1 Local parallel execution

Tests assume one local instance of PTS component, configured for graduate increase of parallelization factor (effectively number of concurrent threads). Platform designated for the run is an Intel Xeon X5355 based workstation, with 4 computing cores and 4 Gbytes of memory. Below results are durations in seconds, of 100 iterations of our TS implementation, depending on the size of the HFS problem ($n$ – task count, $m$ – task stage count, machines in every stage - 5)

|   |   | \multicolumn{5}{c}{thread count} |   |   |   |
|---|---|---|---|---|---|---|
| $n$ | $m$ | 1 | 2 | 3 | 4 | $5^1$ |
| 10 | 2  | 16,2396  | 15,21    | 14,352   | 13,9776  | 14,1648 |
| 10 | 5  | 21,6528  | 18,3768  | 17,5344  | 16,068   | 17,2692 |
| 10 | 8  | 31,7772  | 21,9648  | 19,9992  | 18,0492  | 20,4672 |
| 10 | 10 | 40,716   | 27,5496  | 22,6668  | 21,4032  | 23,4624 |
| 30 | 2  | 54,1476  | 32,3388  | 26,0832  | 24,726   |   |
| 30 | 5  | 180,3828 | 98,9664  | 72,2436  | 60,6684  | 67,5636 |
| 30 | 8  | 427,83   | 224,6644 | 156,177  | 127,678  |   |
| 30 | 10 | 655,681  | 340,001  | 234,5224 | 185,1252 |   |
| 50 | 2  | 177,565  | 99,7152  | 72,7428  | 60,294   |   |
| 50 | 5  | 802,2924 | 423,2904 | 295,5888 | 230,5836 |   |
| 50 | 8  | 1971,137 | 984,329  | 675,769  | 542,577  |   |
| 50 | 10 | 2928,873 | 1482,842 | 1000,248 | 790,129  |   |

**Tab 1: Duration [s] of 100 PTS iterations**

---

[1] Thread count exceeding number of available cores in the system, resulting in decrease of performance, is disregarded in further tests.

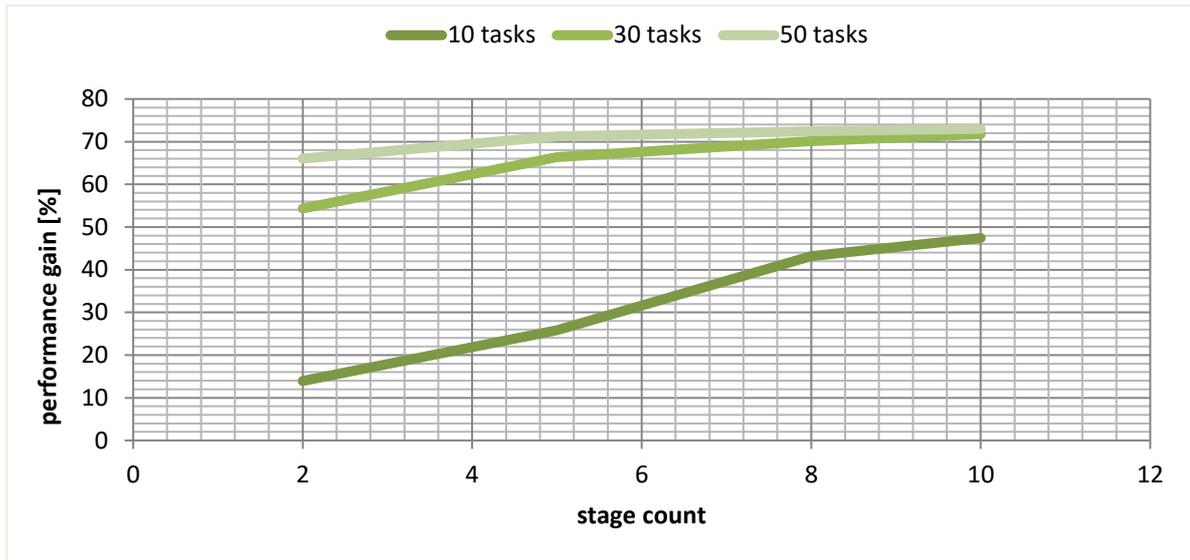

**Fig 5: Duration of 100 PTS iterations**

Above results demonstrate local solution performing proportionally to number of processing units utilized in the system. The performance gain becomes clearly dependent on the size of the problem. It is worth noticing a size threshold of the HFS problem, below which parallelization is hardly beneficial. Given optimization of 10 tasks with 2 stages each, optimization on 4 cores proves to be only 15% faster than single core. Our system was never meant for small problems anyways. Scheduling 10 tasks with 10 stages already gains 50% of time and problems of 30x8 or 50x5 already reach for the *limes* of 75% performance gain, comparing with single threaded execution.

Two more phenomena's are worth mentioning. First of all, parallelizing the processing beyond available processor count, results in performance loss. That loss gives a clue on the cost of context switching between subsequent threads. Secondly, we have observed super linear performance trend, while scheduling 50 tasks with 8 stages.

6.2 Distributed environment

Initial setting consists of 5 hosts in local area network, dedicated to optimization only, and administered by a remote load balancer. All of them are running 32-bit Windows Server 2003 on single-core Intel Celeron 1.7 GHz (*pamięć?*) Focusing on a homogenous environment first, allows for more accurate comparison with the results of local parallelization.

|   |   | Host count | | | | |
|---|---|---|---|---|---|---|
| n | m | 1 | 2 | 3 | 4 | 5 |
| 50 | 2 | 355,47 | 229,375 | 176,15 | 147,42 | 131,81 |
| 50 | 5 | 1658,18 | 897,51 | 625,57 | 546,92 | 406,87 |
| 50 | 8 | 4016 | 2084 | 1449 | 1145 | 922 |
| 50 | 10 | 6110 | 3199,18 | 2168,65 | 1632,45 | 1319,5 |
| 30 | 2 | 100,17 | 100,34 | 88,46 | 82,36 | 81,56 |
| 30 | 5 | 377,25 | 238,78 | 188,09 | 166,62 | 149,125 |
| 30 | 8 | 885,04 | 508,53 | 357,12 | 286,57 | 239,01 |
| 30 | 10 | 1343,15 | 745,06 | 523,09 | 410,57 | 331,23 |

**Table 2: Duration [s] of 100 distributed TS iterations**

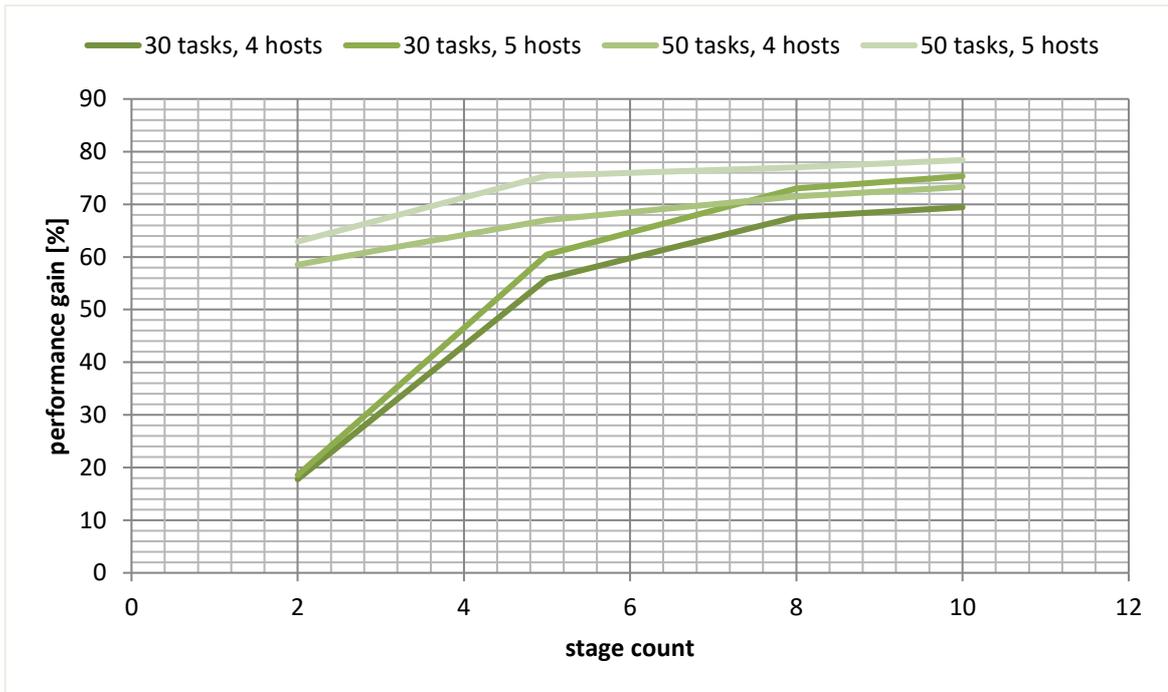

**Fig 5: Duration of 100 distributed TS iterations**

Conducted tests once again indicate complexity of the problem, below which application of the solver makes little sense. 100 iterations of optimizing 20x2 schedule is only 20% faster when ran on 5 machines, than on a single core. Moving towards bigger instances, optimization of 30 tasks with 10 stages is around 75% faster, and that of 50 tasks with 10 stages, almost 79% faster, rendering the scalability and performance of the system above expectations. Promising news is every test, where distribution of the process to distinct hosts turns out slightly more beneficial, than local parallelization. Performance loss due to balancing accuracy and communication bandwidth appears comparable with that of local thread management and scheduling by shared operating system.

Final, stress test, is conducted in a heterogeneous environment, build by extending previous with additional 3 machines in separate subnets, to a set of eight:

- 5 Intel Celeron 1.7 GHz based nodes
- 2 nodes based on Athlon 64 X2 Dual Core 4800+
- 1 server with Intel Core 2 Duo E6750

100 iterations have been run to optimize a schedule of 100 tasks, with 5 stages, each containing 5 machines. Balancer uses the weighted prediction, to load 8 hosts and 11 cores in total. Processing of the problem takes 17 minutes. The same task is finished on a single core Celeron 1.7 in 3:43 hours. Our system finished the optimization in heterogeneous environment over 13 times faster.

## 7. CONCLUSIONS

The developed parallel/distributed algorithm for scheduling multiprocessor tasks in HFS is capable of solving large instances (several hundred of job in 20 stages) in reasonable time period (less than one hour). Moreover, the applied load balancing methods caused that the overhead of computation distribution is very small, and the running time of the proposed algorithm depends almost linearly on the computational power of the machines in the network.